\documentclass[final,3p,times]{elsarticle}

\usepackage{amssymb}
\usepackage{lineno}
\usepackage{hyperref}
\usepackage[figuresright]{rotating}

\newtheorem{proposition}{Proposition}
\newcommand{\proof}{\textbf{Proof\ } }

\journal{Acta Universitatis Palackianae Olomucensis, Facultas rerum naturalium, Mathematica}

\begin{document}
\begin{frontmatter}

\title{A Note on Computing Extreme Tail Probabilities of the Noncentral $T$~Distribution with Large Noncentrality Parameter}
\author[um]{Viktor Witkovsk\'y\corref{cor1}}
\ead{witkovsky@savba.sk}
\cortext[cor1]{Corresponding author. Tel.: +421 2 59104530; Fax: +421 2 54775943.}
\address[um]{Institute of Measurement Science, Slovak Academy of Sciences, Bratislava, Slovakia}

\begin{abstract}
The noncentral $t$-distribution is a generalization of the Student's $t$-distribution.  In this paper we suggest an alternative approach for computing the cumulative distribution function (CDF) of the noncentral $t$-distribution which is based on a direct numerical integration of a well behaved function. With a double-precision arithmetic, the algorithm provides highly precise and fast evaluation of the extreme tail probabilities of the noncentral $t$-distribution, even for large values of the noncentrality parameter $\delta$ and the degrees of freedom $\nu$. The implementation of the algorithm is available at the MATLAB Central, File Exchange:
\href{http://www.mathworks.com/matlabcentral/fileexchange/41790-nctcdfvw}{http://www.mathworks.com/matlabcentral/fileexchange/41790-nctcdfvw}.
\end{abstract}


\end{frontmatter}


{
\section{Introduction}
The noncentral $t$-distribution, originally derived by R.A.~Fisher \cite{Fisher}, is a generalization of the Student's $t$-distribution \cite{Student}. 
Let $Z$ be a random variable with a standard normal distribution, i.e.~$Z\sim N(0,1)$, and let $Q$ be an independent chi-square distributed random variable with $\nu$ degrees of freedom, i.e.~$Q\sim\chi^2_\nu$ with $\nu > 0$. Further, let $\delta$ denote any real constant (the noncentrality parameter), then the probability distribution of the random variable
\begin{equation}\label{eq1}
T = \frac{Z+\delta}{\sqrt{Q/\nu}},
\end{equation}
is called a noncentral $t$-distribution with $\nu$ degrees of freedom  and noncentrality parameter $\delta$, or simply written, $T\sim t_{\nu,\delta}$. If $\delta = 0$ the noncentral $t$-distribution coincides with the central $t$-distribution.

The central as well as the noncentral $t$-distributions belong to the most frequently used distributions in statistics. The cumulative distribution function (CDF) of the noncentral $t$-distribution is used in the power analysis (as a part of statistical inference based on the normal linear models), see e.g.~\cite{Johnson}, including $t$-test as a special case. 
That is, the CDF of the noncentral $t$-distribution is used to evaluate the probability that a $t$-test will correctly reject a false null hypothesis on mean of a normal population $N(\mu,\sigma^2)$, i.e.~the test of the null hypothesis $H_0: \mu \leq \mu_0$ against the alternative $H_A: \mu > \mu_0$ based on small sample from this population, when the population mean $\mu$  is actually greater than $\mu_0$; that is, it gives the power of the $t$-test. 

Broad applicability of the  noncentral $t$-distribution is also in engineering, measurement science and metrology, quality control applications, as well as in financial mathematics. 
An interesting problem with important applications is derivation of the exact confidence interval for the noncentrality parameter $\delta$ based upon a random sample from the normal distribution, or an equivalent problem of derivation of the confidence bounds for the coefficient of variation $c_V = \frac{\sigma}{\mu}$, see e.g.~\cite{Kim}.  The noncentral $t$-distribution is also used for calculating the endpoints of the one-sided tolerance intervals (the tolerance limits) for a normal population. An application of tolerance intervals to manufacturing involves comparing specification limits prescribed by the client with tolerance limits that cover a specified proportion of the population. The required values of the tolerance factors can be computed using the quantiles of the noncentral $t$-distribution. For more details on statistical tolerance intervals and the one-sided tolerance limits see e.g.~\cite{Hahn, Krishnamoorthy, Janiga2006}.

The mathematical expressions for the CDF and the PDF (probability density function) of the noncentral $t$ are rather complicated. 
There  is not known closed form (analytical) expression of the CDF. So, it is typically given as a series expansion in terms of incomplete beta functions and/or other special functions. A comprehensive list of useful mathematical expressions and alternative representations can be found e.g.~in \cite{Johnson}. 

Algorithms for numerical computation of the CDF, PDF, and the quantiles of the noncentral $t$-distribution are available in numerical libraries (e.g.~BOOST C++ Libraries \cite{boost}), and different implementations are typically available also in standard statistical packages and/or programming environments (e.g.~MATLAB \cite{Matlab}, \textsl{R} \cite{R}, SAS \cite{SAS}, and Mathematica \cite{Mathematica}).

However,  applicability of the currently used implementations 
could be limited with respect to the speed and/or precision. This is true especially if the interest is in highly precise evaluation of the extreme tail probabilities of $t$-distributions with large values of the noncentrality parameter $\delta$ and/or the degrees of freedom $\nu$. 

The aim of this paper is to provide a brief overview of the basic properties of the noncentral $t$-distribution, useful for numerical evaluation of the CDF and PDF (consequently also for evaluation of the quantiles of the distribution), and comparisons of the current implementations, especially if used for evaluation of the extreme tail probabilities with large values of the noncentrality parameter. 

As an alternative, we have developed a new algorithm based on direct numerical integration (by using standard Gauss-Kronod quadrature) of a well behaved function which leads to highly precise and fast evaluation of the CDF for any combination of the input parameters $x$, $\nu$, and $\delta$.

\section{The noncentral $t$-distribution}
Here we shall briefly summarize some of the known properties of the noncentral $t$-distribution useful for numerical evaluation of the CDF, for more details see e.g.~\cite{Abramowitz}  and \cite{Johnson}.

\begin{proposition}
Let $F_{t_{\nu,\delta}}(x) =  \Pr(T\leq x)$ be the CDF of the  random variable $T\sim t_{\nu,\delta}$, defined by (\ref{eq1}) for any real $x$, $\delta$ and $\nu>0$.
Then the following properties hold true:

\begin{enumerate}
\item[1.] If $x = 0$, then 
\begin{equation}
		F_{t_{\nu,\delta}}(x) = \Phi(-\delta),
\end{equation}
where $\Phi(\cdot)$ is the CDF of standard normal distribution.
	
\item[2a.] If $x>0$, then 
\begin{equation}\label{2a}
		F_{t_{\nu,\delta}}(x) = \int_{0}^\infty\Phi\left( x\,\sqrt{\frac{q}{\nu}}-\delta\right)f_{\chi^2_\nu}(q)\,dq,
\end{equation}
where $f_{\chi^2_\nu}(\cdot)$ is the PDF of chi-square distribution with $\nu$ degrees of freedom.

\item[2b.] If $x>0$, then 
	\begin{eqnarray}\label{2b}
	F_{t_{\nu,\delta}}(x) &=& \Phi(-\delta) + \int_{-\delta}^\infty\left(1-F_{\chi^2_\nu}\left(\frac{\nu(z+\delta)^2}{x^2}\right)\right)\phi(z)\,dz\cr
	&=& \Phi(-\delta) + \int_{-\delta}^\infty \Gamma_{u}\left( \frac{\nu}{2},\frac{\nu(z+\delta)^2}{2 x^2}\right)\phi(z)\,dz\cr
	&=& 1 -\int_{-\delta}^\infty \Gamma_{l}\left( \frac{\nu}{2},\frac{\nu(z+\delta)^2}{2 x^2}\right)\phi(z)\,dz,
\end{eqnarray}
where $F_{\chi^2_\nu}(\cdot)$ is the CDF of chi-squared distribution with $\nu$ degrees of freedom, $\phi(\cdot)$ is the PDF of standard normal distribution, 
$\Gamma_{u}(\cdot,\cdot)$  (resp.~$\Gamma_{l}(\cdot,\cdot)$) denotes the upper (lower) regularized incomplete gamma function, i.e.~$\Gamma_{u}(a,x) = \frac{1}{\Gamma(a)}\int_{x}^\infty e^{-t} t^{a-1}\, dt$, $\Gamma_{l}(a,x) = 1-\Gamma_{u}(a,x)$, and $\Gamma(\cdot)$ is the  gamma function.	

\item[2c.] If $x>0$, then 
\begin{equation}\label{2c}
	F_{t_{\nu,\delta}}(x) = \Phi(-\delta) + \frac{1}{2}\sum_{i=0}^\infty \left\{P_i I_y\left(i+\frac{1}{2},\frac{\nu}{2}\right) + \frac{\delta}{\sqrt{2}}\,Q_i I_y\left(i+1,\frac{\nu}{2}\right) \right\},
\end{equation}
where 
\begin{equation}
	P_i = \frac{(\delta^2/2)^i}{i!} e^{-\frac{\delta^2}{2}},\quad Q_i = \frac{(\delta^2/2)^i}{\Gamma(i+3/2)} e^{-\frac{\delta^2}{2}}, \quad y = \frac{x^2}{\nu + x^2}, 
\end{equation}
and $I_y(a,b)$ is the incomplete beta function.

\item[3.] If $x<0$, then 
\begin{equation}
		F_{t_{\nu,\delta}}(x) = 1 - 	F_{t_{\nu,-\delta}}(-x).
\end{equation}
\end{enumerate}
\end{proposition}

\noindent
\proof
The CDF representation given in (\ref{2c})~has been proved by Guenther in \cite{Guenther}. The other properties can be derived directly from the definition of the noncentral $t$-variable (\ref{eq1}). In particular, 
\begin{equation}
	F_{t_{\nu,\delta}}(0) = \Pr(T\leq 0) = \Pr\left(\frac{Z+\delta}{\sqrt{Q/\nu}}\leq0 \right) = \Pr\left(Z\leq -\delta \right) = \Phi(-\delta).
\end{equation}
If $x>0$, then we get
\begin{eqnarray}
	F_{t_{\nu,\delta}}(x) =  \Pr(T\leq x) &=& \Phi(-\delta) + \Pr(0< T\leq x)\cr
	&=& \Phi(-\delta) + \Pr \left(-\delta < Z \leq x\sqrt{\frac{Q}{\nu}}-\delta \right)\cr
	&=& \Phi(-\delta) + E_{\{Q\}}\left(  \Phi\left(  x\sqrt{\frac{Q}{\nu}}-\delta\right)\right) - \Phi(-\delta)\cr
	&=& E_{\{Q\}}\left( \Phi\left(  x\sqrt{\frac{Q}{\nu}}-\delta\right) \right) = \int_{0}^\infty\Phi\left( x\,\sqrt{\frac{q}{\nu}}-\delta\right)f_{\chi^2_\nu}(q)\,dq,
\end{eqnarray}
where $Q\sim\chi^2_\nu$, $f_{\chi^2_\nu}(\cdot)$  is the PDF of $\chi^2_\nu$-distribution, and $E_{\{Q\}}(\cdot)$ denotes the expectation operator (with respect to the distribution of a random variable $Q$). 
Similarly, we get
\begin{eqnarray}
	F_{t_{\nu,\delta}}(x) &=& \Phi(-\delta) + \Pr \left(-\delta < Z \leq x\sqrt{\frac{Q}{\nu}}-\delta \right)\cr
	&=& \Phi(-\delta) + \Pr \left(Z \leq x\sqrt{\frac{Q}{\nu}}-\delta \,\Big|\, Z > -\delta \right)\cr
	&=&\Phi(-\delta) +  \Pr \left(\frac{\nu(Z+\delta)^2}{x^2}  \leq Q \,\Big|\, Z > -\delta \right)\cr
	&=& \Phi(-\delta) + \left(1- \Pr \left(Q\leq\frac{\nu(Z+\delta)^2}{x^2}  \,\Big|\, Z > -\delta \right) \right)\cr
	&=& \Phi(-\delta) + E_{\{Z>-\delta\}} \left( 1-F_{\chi^2_\nu}\left( \frac{\nu(Z+\delta)^2}{x^2}\right)\right)\cr
	&=& \Phi(-\delta) + \int_{-\delta}^\infty \Gamma_{u}\left( \frac{\nu}{2},\frac{\nu(z+\delta)^2}{2 x^2}\right)\phi(z)\,dz\cr
  &=& 1 -\int_{-\delta}^\infty \Gamma_{l}\left( \frac{\nu}{2},\frac{\nu(z+\delta)^2}{2 x^2}\right)\phi(z)\,dz
\end{eqnarray}
where $F_{\chi^2_\nu}(q) = \Gamma_{l}\left( \frac{\nu}{2},\frac{q}{2}\right) = 1- \Gamma_{u}\left( \frac{\nu}{2},\frac{q}{2}\right)$ is the CDF of chi-squared distribution with $\nu$ degrees of freedom, and $\Gamma_{l}\left( \frac{\nu}{2},\frac{q}{2}\right)$ (resp.~$\Gamma_{u}\left( \frac{\nu}{2},\frac{q}{2}\right)$) denotes the lower (upper) regularized incomplete gamma function, and $\phi(\cdot)$ is the PDF of standard normal distribution. Note that the representation holds true also for noninteger degrees of freedom, $\nu>0$. Finally, if $x<0$, we get
\begin{eqnarray}
F_{t_{\nu,\delta}}(x) &=& \Pr(T\leq x) = \Pr\left(\frac{Z+\delta}{\sqrt{Q/\nu}}\leq x \right) = \Pr\left(\frac{-Z-\delta}{\sqrt{Q/\nu}}> -x \right)\cr
&=& 1-	\Pr\left(\frac{Z-\delta}{\sqrt{Q/\nu}}\leq -x \right)=1- F_{t_{\nu,-\delta}}(-x),
\end{eqnarray}
by using the symmetry of the distribution of the random variable $Z\sim N(0,1)$. \hfill $\square$

The CDF $	F_{t_{\nu,\delta}}(\cdot)$ can be directly used for computing the PDF of the noncentral $t$ distribution, $f_{t_{\nu,\delta}}(\cdot)$, defined by $f_{t_{\nu,\delta}}(x) = \partial F_{t_{\nu,\delta}}(x)/ \partial x$. In particular, the following holds true:

\begin{itemize}
\item If $x = 0$, then 
\begin{equation}
		f_{t_{\nu,\delta}}(x) = \frac{\Gamma\left(\frac{\nu+1}{2}\right)}{\sqrt{\pi \nu}\, \Gamma\left(\frac{\nu}{2}\right)}\, e^{-\frac{\delta^2}{2}}.
\end{equation}

\item If $x \neq 0$, then 
\begin{equation}
		f_{t_{\nu,\delta}}(x) = \frac{\nu}{x}\left\{F_{t_{\nu+2,\delta}}\left( x\, \sqrt{1+\frac{2}{\nu}}\right) -  F_{t_{\nu,\delta}}(x) \right\}.
\end{equation}
\end{itemize}

Based on that, the quantiles of the noncentral $t$-distribution, say $x_p$, can be calculated for any given $p\in (0,1)$. In general, for any fixed combination of the three parameters (from the set $\nu$, $\delta$, $x$, $p$), the remaining parameter can be calculated either directly, or via the usual root-finding techniques, by solving the equation $F_{t_{\nu,\delta}}(x) = p$.

\section{Standard implementations for computing CDF of the noncentral $t$-distribution}

The standard algorithm implementations for computing the CDF of the noncentral $t$-distribution are based on its representation (\ref{2c}), which was originally derived by Guenther \cite{Guenther} and later implemented by Lenth \cite{Lenth}. 

Due to the recurrence properties of the incomplete beta function, the algorithm requires only two evaluations of $I_y(a,b)$ and the rest is based on simple arithmetic operations. For most typical values of the input arguments $x$, $\nu$, and $\delta$, the algorithm is extremely fast and accurate.

The \textsl{R} implementation (function \textsl{pt} in \cite{R}) is based on \textsl{C} version of the Lenth's algorithm with a restricted range of the noncentrality parameter, $|\delta|\leq 37.62$. Otherwise, the result is based on  normal approximation, 

\begin{equation}
F_{t_{\nu,\delta}}(x) = \Phi(z), \quad \mathrm{ where } \quad z = \frac{x(1-\frac{1}{4\nu})-\delta}{\sqrt{1+\frac{x^2}{2\nu}}},   
\end{equation}
see \cite{Abramowitz}, eqn.~(26.7.10), p.~949, which can be rather poor for small $\nu$.

Algorithm based on (\ref{2c})~have been implemented also in MATLAB (function \textsl{nctcdf} in \cite{Matlab}) and in the BOOST C++ Libraries (function \textsl{non\_central\_t} in \cite{boost}, see also \cite{DISTEXPLORER}). The BOOST implementation is based on strategies suggested by Benton and Krishnamoorthy \cite{Benton}. 

The BOOST function \textsl{non\_central\_t} has been tested for wide range of input parameters and compared with test data computed by arbitrary precision interval arithmetic (believed to be accurate to at least 50 decimal places, as declared in \cite{boost}, and confirmed by a large test data set, kindly provided by J.~Maddock [personal communication]). As the complexity of the algorithm based on the series expansion as given in (\ref{2c}) is dependent upon $\delta^2$, consequently, the time taken to evaluate the CDF increases rapidly for large noncentrality parameter, $|\delta| > 500$, likewise the accuracy decreases rapidly for very large $\delta$, see \cite{boost}.

Moreover, unlike the \textsl{R} and MATLAB implementations, which compute correctly only the lower tail of the distribution,  the BOOST algorithm computes also the upper tail of the distribution (which is important for correct evaluation of the extreme tail probabilities).

As presented in \cite{SAS}, SAS implementation (function \textsl{probt}) is based on numerical integration of the representation (\ref{2a}). For most typical values of the input arguments $x$, $\nu$, and $\delta$, the algorithm is  fast and accurate (for most cases, typically all 14 reported significant digits are correct). However, for more extreme input arguments the algorithm may fail to converge to the prescribed accuracy, and in such case no output is provided by the function  \textsl{probt}.

Implementation in Mathematica (function \textsl{CDF[NoncentralStudentTDistribution[$\nu$, $\delta$], $x$]} in \cite{Mathematica}) is based on numerical integration (computed using Mathematica's high-precision arithmetic) of the noncentral $t$ PDF function (which is given as an analytical function expressible by using the Hermite polynomials). The computational complexity of this algorithm quickly grows with large $\nu$ and $\delta$ and in such cases fail to converge.

\section{Algorithm \textsl{nctcdfvw} based on direct numerical integration}

As an alternative, here we suggest a new algorithm based on direct numerical integration (by using standard Gauss-Kronod quadrature) of a well behaved function, based on expression \emph{2b}, which leads to highly precise and fast evaluation of the CDF for any combination of the input parameters $x$, $\nu$, and $\delta$. 

The algorithm has been implemented in MATLAB and its current version is available at the MATLAB Central, File Exchange: \url{http://www.mathworks.com/matlabcentral/fileexchange/41790-nctcdfvw}. 

The algorithm computes the lower tail, i.e. $\Pr(T\leq x)= \Phi(-\delta) + \int_{-\delta}^\infty \Gamma_{u}\left( \frac{\nu}{2},\frac{\nu(z+\delta)^2}{2 x^2}\right)\phi(z)\,dz$, if $0< x \leq \delta$, otherwise, for $x > 0$ and such that $x > \delta$,  it computes the upper tail of the distribution, $\Pr(T> x) = \int_{-\delta}^\infty \Gamma_{l}\left( \frac{\nu}{2},\frac{\nu(z+\delta)^2}{2 x^2}\right)\phi(z)\,dz$.  

Notice that in a double-precision arithmetic the integration range $(-\delta,\infty)$ 
can be reduced to the limits $[A_0,B_0]$ given by
\begin{equation}
	[A_0,B_0] = [\max(-\delta,\Phi^{-1}(r_{\varepsilon_0})),-\Phi^{-1}(r_{\varepsilon_0})], 
\end{equation}
where $r_{\varepsilon_0}$ is the minimum real non-zero number (the smallest positive normalized floating point number in IEEE double precision), i.e.~$r_{\varepsilon_0} =   2.2251\times 10^{-308}$.  
So, $\Phi^{-1}(r_{\varepsilon_0}) =  -37.5194$. 

The most important part of the algorithm is the method for subsequent (significant) reduction of the integration range $[A_0,B_0]$. For simplicity,  we shall illustrate this only for the case of computing the lower tail of the distribution. 

Notice that $\Gamma_{u}\left( \frac{\nu}{2},\frac{\nu(z+\delta)^2}{2 x^2}\right)$ approaches the value $1$ for small values of $z$ ($z\rightarrow-\delta$) and $0$ for large values of $z$ ($z\rightarrow +\infty$). If $\Gamma_{u}\left( \frac{\nu}{2},\frac{\nu(z+\delta)^2}{2 x^2}\right)> 1-\varepsilon_{R}$ for $ A_0\leq z \leq A_1$, where 
\begin{equation}\label{A1limit}
	A_1 = \sqrt{\frac{x^2 	q_{\varepsilon_{R}} }{\nu}}-\delta,
\end{equation}
with $q_{\varepsilon_{R}}$ being the $\varepsilon_{R}$-quantile of the $\chi^2_\nu$-distribution (for $q_{\varepsilon_{R}}\approx 0$ set $q_{\varepsilon_{R}} = 0$), where $\varepsilon_{R}$ is the required relative tolerance bound (in double-precision arithmetic we set $\varepsilon_{R} = 10^{-16}$), then the integration range can be further reduced to
\begin{equation}
	[A_1,B_1] = [\max(A_0,A_1),B_0],
\end{equation}
and the CDF is approximated by
\begin{equation}
	\Pr(T\leq x)\approx \Phi(A_1) + \int_{A_1}^{B_1} \Gamma_{u}\left( \frac{\nu}{2},\frac{\nu(z+\delta)^2}{2 x^2}\right)\phi(z)\,dz
\end{equation}

Further reduction of the integration range $[A_1,B_1]$ is possible since the integrand function 
\begin{equation}\label{integrand}
	g(z) = \Gamma_{u}\left( \frac{\nu}{2},\frac{\nu(z+\delta)^2}{2 x^2}\right)\phi(z) 
\end{equation}
(typically) quickly fades out from its maximum value. 
For illustration, Figure~\ref{fig} presents a typical graph of the integrand function $g(z)$, together with the optimally selected integration limits.

\begin{figure}[t]
\begin{center}
\includegraphics[width=.9\columnwidth]{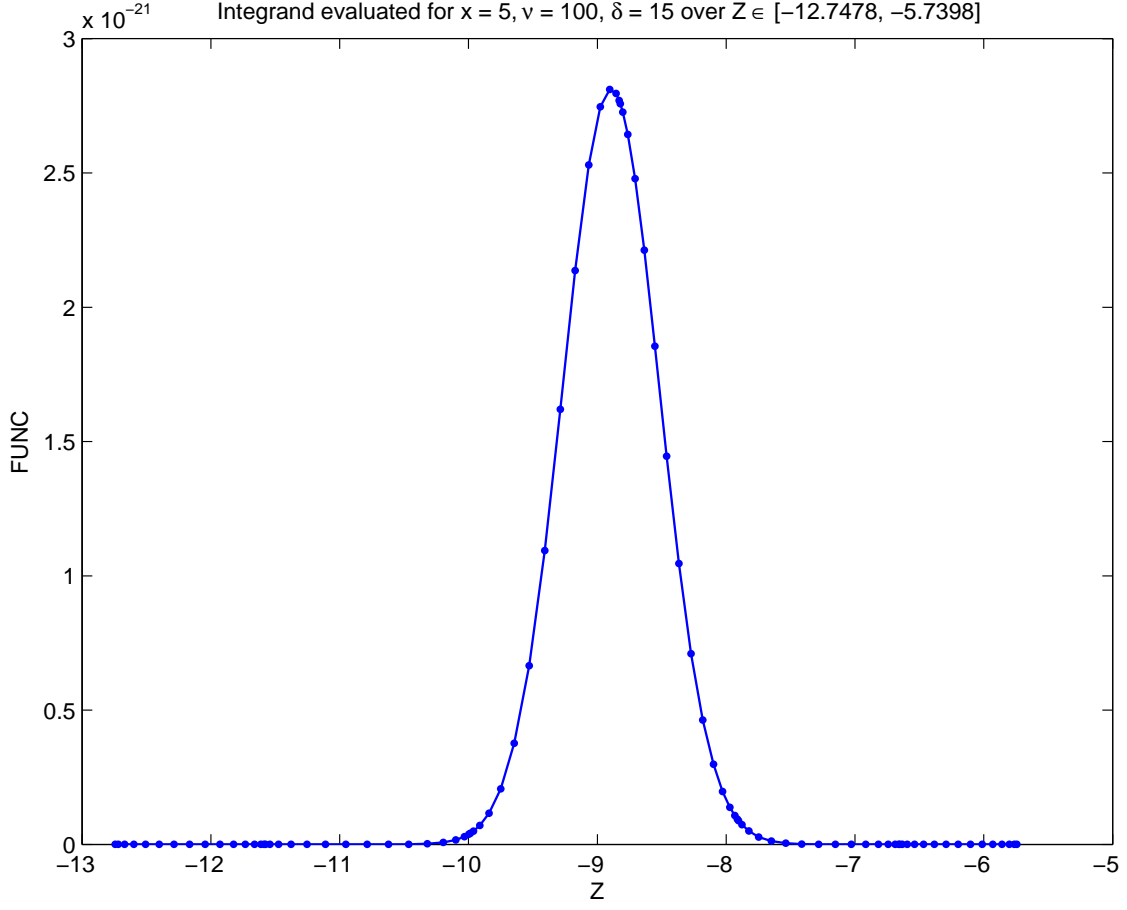}
\end{center}
\caption{The integrand function $g(z)$ evaluated at the Gauss-Kronod nodes ($90 = 6 \ \mathrm{subintervals}\times 15\ \mathrm{nodes}$) and the selected integration limits $[A,B]$ used for computing the CDF of the noncentral $t$-distribution with the input parameters $x = 5$, $\nu=100$ and $\delta = 15$.  The value of CDF computed by the algorithm \textsl{nctcdfvw} is $\mathrm{CDF} =   2.640405806735035\times 10^{-21}$.  For comparison, the standard MATLAB function \textsl{nctcdf} (Statistics Toolbox) returns $\mathrm{CDF} =  4.542511227039881\times 10^{-43}$, the \textsl{R} function \textsl{pt} returns  $\mathrm{CDF} =  2.35515251660662\times 10^{-21}$, the SAS function \textsl{probt} returns $\mathrm{CDF} =  2.6404058074408\times 10^{-21}$, and the BOOST function \textsl{non\_central\_t} returns $\mathrm{CDF} =  2.64040580673507\times 10^{-21}$.
}\label{fig}
\end{figure}

Based on that, further reduction of the integration range is given, and the final integration limits are given by
\begin{equation}\label{limits}
	[A,B] = [\max(A_1,A_2),\min(B_1,B_2)], 
\end{equation}
where the limits $A_2$ and $B_2$ are given as the two possible solutions to the equation
\begin{equation}\label{ablimits}
	g(A_2) = g(B_2) = \varepsilon_{A},
\end{equation}
where $\varepsilon_{A}$ denotes the required absolute tolerance bound (which should be properly estimated, see bellow).

The integrand function $g(z)$ defined by (\ref{integrand}) and its modus can be effectively  estimated based on the results and efficient approximations of the CDF and the quantiles of the chi-square distribution, as suggested by Inglot in \cite{Inglot}. 

In particular, let us denote $q = \frac{\nu(z+\delta)^2}{x^2}$ and $h(z) = \log(g(z))$. 
Then, by using the lower bound for tails of the $\chi^2_\nu$-distribution, as derived 
in \cite{Inglot} and \cite{Ledwina}, i.e.~$\frac{1}{2}\mathcal{E}_\nu(q)\leq \Pr(\chi^2_\nu > q) \leq \frac{1}{\sqrt{\pi}}\frac{q}{q-\nu+2} \mathcal{E}_\nu(q)$, where 
\[
\mathcal{E}_\nu(q) = \exp\left\{-\frac{1}{2}\left(q-\nu-(\nu-2)\log\left( \frac{q}{\nu}\right)+\log(\nu) \right) \right\},
\] 
we get
\begin{eqnarray}
	h(z) &\approx& \log\left( \frac{1}{2}\mathcal{E}_\nu(q)\phi(z)\right) \cr
	&\approx& -\log(2) -\frac{1}{2}\left(q - \nu -(\nu-2)\log\left(\frac{q}{\nu}\right)+\log(\nu)+\log(2\pi)+z^2 \right),
\end{eqnarray}
which holds true for all $q$ (and consequently for all $z$) and $\nu>2$. For practical purposes, the algorithm sets $\nu-2 \equiv 1$ if $\nu\leq 2$.

By solving the equation $\frac{\partial h(z)}{\partial z} = 0$ we get the estimate of the mode (modus), say  $z_{mod}$, of the integrand function $g(z)$ as
\begin{equation}
	z_{mod} = \frac{-\delta(x^2+2\nu)+x\sqrt{4\nu(\nu-2)+x^2(\delta^2+4(\nu-2))}}{2\left(x^2+\nu\right)}.
\end{equation}
Based on that, we can estimate the maximum value of the integrand function by 
\begin{equation}
	g_{max}\approx g(z_{mod}) \approx \exp\left( h(z_{mod})\right),
\end{equation}
and also the required absolute tolerance bound $\varepsilon_{A}$, by solving
\begin{equation}
\log\left(\varepsilon_{A}\right) \approx   h(z_{mod}) + \log\left(\varepsilon_{R}\right).
\end{equation}
Consequently, the limits $A_2$ and $B_2$ defined by (\ref{ablimits}) can be estimated by solving the approximate equation
\begin{equation}
	h(A_2) = h(B_2) = \log(\varepsilon_{A}).
\end{equation}
The algorithm \textsl{nctcdfvw} finds the limits $A_2$ and $B_2$ by solving quadratic equation which results from the quadratic approximation (expansion) of the function $h(\cdot)$ around $z_{mod}$.

The quantiles $q_{\varepsilon_{R}}$ used in (\ref{A1limit}) are estimated by using  efficient approximation proposed in \cite{Inglot}, eqn.~(A.3),
\begin{eqnarray}
	q_{\varepsilon_{R}} &\approx& \nu +2\varepsilon_{R} + 1.62\sqrt{\nu \varepsilon_{R}} + 0.63012\sqrt{\nu}\log(\varepsilon_{R}) \cr
	& & -1.12032\sqrt{\nu} - 2.48\sqrt{\varepsilon_{R}} - 0.65381\log(\varepsilon_{R}) -0.22872.
\end{eqnarray}

Given the integration limits $[A,B]$, the algorithm evaluates the CDF by using the approximation
\begin{equation}
	F_{t_{\nu,\delta}}(x) \approx \Phi(A) + \int_{A}^{B} g(z)\,dz
\end{equation}
The integral $\int_{A}^{B} g(z)\,dz$ can be evaluated by using the standard (adaptive) Gauss-Kronod quadrature which allows to estimate the integration error. 

In order to speed-up the computation by using evaluation of the vectorized functions, the MATLAB version of the algorithm \textsl{nctcdfvw} uses the non-adaptive version of the (G7,K15)-Gauss-Kronod quadrature over fixed (prespecified) number of sub-intervals of $[A,B]$. The default number of sub-intervals is set to $n_{subs} = 16$ (a rather conservative choice based on a detailed and extensive testing, in order to ensure that the relative precision to be better than (or equal to) $10^{-14}$, in most cases), but for most typical values of the input parameters division to 6 sub-intervals (which requires $90 = 6\times 15$ evaluations of the integrand function $g(z)$) is sufficient to achieve the relative error less than $10^{-12}$.

\section{Accuracy comparisons}
In order to illustrate and compare the accuracy of the standard algorithms/implementations for computing the noncentral $t$ distribution (and to compare it with the suggested algorithm based on the expression (\ref{2b})), Table~\ref{tab} presents the CDF values of the noncentral $t$ distribution  for several (rather extreme) combinations of input parameters $x$, $\nu$, and $\delta$, computed by different algorithms/implementations. In particular,
\begin{itemize}
	\item \textbf{MATLAB} function \textbf{\textsl{nctcdfvw}} (based on non-adaptive Gauss-Kronod quadrature, here with integration limits $[A,B]$ divided into 32 subintervals),  
	\item \textbf{MATLAB} function \textbf{\textsl{nctcdf}} (Statistics Toolbox),
	\item \textbf{R} function \textbf{\textsl{pt}},
	\item \textbf{SAS} function \textbf{\textsl{probt}},
	\item \textbf{BOOST} C++ Libraries function \textbf{\textsl{non\_central\_t}}, as implemented in \textbf{\textsl{DistExplorer}},
	\item \textbf{MATHEMATICA} function  \textbf{\textsl{CDF[NoncentralStudentTDistribution]}}.
\end{itemize}

The 'true' values of the CDF have been computed by a version of the MATLAB algorithm \textsl{nctcdfvw} (modified for quadruple-precision computation by using the Multiprecision Computing Toolbox for MATLAB \cite{Holoborodko}). 

All computations have been realized on standard PC under 32-bit Windows XP operating system. For detailed comparisons, the results are presented with 18 significant digits.  Notice however, that the double-precision arithmetic (used by the presented algorithms, except Mathematica) returns only 16 significant digits. 

The differences (with respect to the exact values) are emphasized by underlining the affected digits. The 'NA' value is displayed if the algorithm did not converge. A symbol '*' is displayed for cases when Mathematica warning message  '\textsl{NIntegrate} failed to converge to prescribed accuracy' has been issued.

\begin{sidewaystable}
{\renewcommand{\baselinestretch}{1.1} 
\small
\centering
\begin{tabular}{rrrrrrrr}
\hline

 {\bf Example} &    {\bf $x$} &   {\bf $\nu$} & {\bf $\delta$} & \multicolumn{1}{c}{\bf TRUE CDF} &  \multicolumn{1}{c}{\bf NCTCDFVW} & \multicolumn{1}{c}{\bf MATLAB} &  \multicolumn{1}{c}{\bf R } \\
\hline

         1 &     1\  &       1\  &       0\  & 7.50000000000000000E-001 &  7.50000000000000000E-001 & 7.50000000000000000E-001 & 7.50000000000000\underline{200}E-001 \\

         2 &   -35\  &       1\  &       0\  & 9.09209467564843408E-003 &      9.09209467564843\underline{700}E-003 & 9.09209467564843\underline{500}E-003 & 9.09209467564843\underline{300}E-003 \\

         3 &   -35\  &       1\  &       1\  & 1.89903487263458750E-003 &      1.89903487263458\underline{800}E-003 & 1.8990348726345\underline{4200}E-003 & 1.899034872\underline{74889500}E-003 \\

         4 &    -5\  &       1\  &       5\  & 8.52042451613777143E-009 &      8.52042451613777\underline{000}E-009 &  8.52042\underline{303378652800}E-009 & 8.520\underline{61943223958500}E-009 \\

         5 &   -15\  &       1\  &      15\  & 1.29043391190105994E-053 &      1.2904339119010\underline{6200}E-053 &          \underline{0} & \underline{8.16013923099490100E-014} \\

         6 &   -35\  &       1\  &      35\  & 7.31501102529248499E-272 &     7.3150110252924\underline{7900}E-272 &         \underline{0} & \underline{5.72875080706580800E-014} \\

         7 &     1\  &      10\  &       5\  & 4.34725285650591657E-005 &      4.34725285650591\underline{700}E-005 &  4.3472528565059\underline{2700}E-005 & 4.3472528\underline{4300420300}E-005 \\

         8 &     1\  &      10\  &      10\  & 7.95914542988750673E-019 &      7.95914542988750\underline{700}E-019 &  \underline{7.61985302416059300E-024} & 7.9\underline{4724820780974200}E-019 \\

         9 &     1\  &      10\  &      15\  & 1.41346486009205976E-042 &      1.413464860092059\underline{00}E-042 &  \underline{3.67096619931285900E-051} & \underline{9.87042896134168300E-043} \\

        10 &     1\  &      10\  &      35\  & 1.69061467860900429E-237 &     1.69061467860900\underline{800}E-237 & \underline{1.12491070647255300E-268} & \underline{4.51922957217374300E-250} \\

        11 &   150\  &      10\  &     200\  & 5.88999020094520836E-002 &      5.88999020094520\underline{500}E-002 & 5.889990200\underline{72176300}E-002 & 5.4599520\underline{1321134400}E-002 \\

        12 &   150\  &      10\  &     500\  & 3.25241635439258347E-019 &      3.2524163543925\underline{7900}E-019 &  3.252416354\underline{69807400}E-019 & \underline{2.762470663506697e00-026} \\

        13 &    50\  &     100\  &      75\  & 4.99615060338271916E-011 &      4.99615060338271\underline{200}E-011 & 4.9961506033\underline{7484300}E-011 &          \underline{0} \\

        14 &   500\  &     100\  &     510\  & 3.71160937464178059E-001 &      3.7116093746417\underline{7900}E-001 & 3.71160937\underline{385911300}E-001 & 3.7\underline{5215597825695000}E-001 \\

        15 &     1\  &    1000\  &      10\  & 1.14935521338266224E-019 &      1.149355213382662\underline{00}E-019 &  \underline{7.61985302416059300E-024} & 1.1493\underline{2401536150700}E-019 \\

        16 &   100\  &    1000\  &     105\  & 2.05403544901854621E-002 &      2.05403544901854\underline{000}E-002 & 2.0540354490\underline{0927000}E-002 & 2.0\underline{1116859042299000}E-002 \\

        17 &  1000\  &    1000\  &    1010\  & 3.22438286661716843E-001 &      3.22438286661716\underline{400}E-001 &  3.22438286\underline{530340600}E-001 & 3.2\underline{3499129176421900}E-001 \\

\hline

{\bf Example} &    {\bf $x$} &   {\bf $\nu$} & {\bf $\delta$} &  \multicolumn{1}{c}{\bf TRUE CDF} &  \multicolumn{1}{c}{\bf SAS} & \multicolumn{1}{c}{\bf BOOST} &  \multicolumn{1}{c}{\bf MATHEMATICA} \\
\hline

         1 &     1\  &       1\  &       0\  & 7.50000000000000000E-001 & 7.50000000000000000E-001 & 7.50000000000000000E-001 & 7.50000000000000000E-001 \\

         2 &   -35\  &       1\  &       0\  & 9.09209467564843408E-003 & 9.0920946756484\underline{0000}E-003 & 9.09209467564843\underline{000}E-003 & 9.09209467564843408E-003 \\

         3 &   -35\  &       1\  &       1\  & 1.89903487263458750E-003 &         NA & 1.8990348726345\underline{9000}E-003 & 1.89903487263458750E-003 \\

         4 &    -5\  &       1\  &       5\  & 8.52042451613777143E-009 &  8.5204245161377\underline{0000}E-009 & 8.520424\underline{42641561000}E-009 & 8.52042451613777143E-009 \\

         5 &   -15\  &       1\  &      15\  & 1.29043391190105994E-053 & 1.2904339119010\underline{0000}E-053 & \underline{2.86650837419345000E-016} & 1.29043391190105994E-053 \\

         6 &   -35\  &       1\  &      35\  & 7.31501102529248499E-272 &         NA &\underline{ -3.77352003162056000E-017} & 7.31501102529248499E-272 \\

         7 &     1\  &      10\  &       5\  & 4.34725285650591657E-005 & 4.3472528565059\underline{0000}E-005 & 4.34725285650591\underline{000}E-005 & 4.34725285650591657E-005 \\

         8 &     1\  &      10\  &      10\  & 7.95914542988750673E-019 & 7.95914542988750\underline{000}E-019 & 7.9591454298875\underline{2000}E-019 & 7.95914542988750673E-019 \\

         9 &     1\  &      10\  &      15\  & 1.41346486009205976E-042 & 1.4134648600920\underline{0000}E-042 & 1.4134648600920\underline{7000}E-042 & 1.41346486009205976E-042 \\

        10 &     1\  &      10\  &      35\  & 1.69061467860900429E-237 & 1.69061467860\underline{890000}E-237 & \underline{1.12491070647253000E-268} & 1.69061467860900429E-237 \\

        11 &   150\  &      10\  &     200\  & 5.88999020094520836E-002 & 5.88999020094520\underline{000}E-002 & 5.88999020094\underline{668000}E-002 &         NA \\

        12 &   150\  &      10\  &     500\  & 3.25241635439258347E-019 &         NA & 3.25241635439\underline{664000}E-019 &         NA \\

        13 &    50\  &     100\  &      75\  & 4.99615060338271916E-011 & 4.996150603382\underline{60000}E-011 & 4.9961506033827\underline{2000}E-011 & *4.99615060338271916E-011 \\

        14 &   500\  &     100\  &     510\  & 3.71160937464178059E-001 &         NA & 3.711609374641\underline{69000}E-001 &         NA \\

        15 &     1\  &    1000\  &      10\  & 1.14935521338266224E-019 & 1.1493552133826\underline{0000}E-019 & 1.1493552133826\underline{9000}E-019 & *1.14935521338266224E-019 \\

        16 &   100\  &    1000\  &     105\  & 2.05403544901854621E-002 & 2.0540354490185\underline{0000}E-002 & 2.054035449018\underline{43000}E-002 & *2.05403544901854621E-002 \\

        17 &  1000\  &    1000\  &    1010\  & 3.22438286661716843E-001 &         NA & 3.224382866616\underline{95000}E-001 &         NA \\

\hline
\end{tabular}
\caption{CDF values of the noncentral $t$-distribution computed by different algorithms/implementations for selected combinations of the input parameters. 
}\label{tab}
}
\end{sidewaystable}

\section{Conclusions}
It seems that (currently) there is no implementation of the algorithm for computing CDF of the noncentral $t$-distribution which is uniformly efficient (reasonably fast) and accurate for all input parameters $x$, $\nu$, and $\delta$ in double-precision arithmetic. 

According to our present study, this goal is best satisfied by the SAS and  BOOST implementations, if we restrict to the typical (most frequently used) values of the input parameters. 
However, when the output of such algorithm is supposed to be used subsequently for further computations, as e.g.~computing the PDF or quantiles of the distribution (and/or the noncentrality parameter $\delta$, or the degrees of freedom $\nu$, for given values $x$ and the CDF/PDF), the possible inaccuracy, slow evaluation or failure to converge, can be critical.

Here we have suggested a new algorithm based on numerical quadrature of a well behaved function which is reasonably precise and fast in double-precision arithmetic for all input parameters $x$, $\nu$ and $\delta$. Precision of the MATLAB version of the algorithm was tested for wide range of input parameters (not presented here). In most of the tested cases the relative error was bellow $10^{-14}$.
}

\section{Acknowledgements}
The author wishes to thank two anonymous referees for identifying the errors in  original version of the manuscript and for their constructive comments which helped to improve presentation of the paper. The work was supported by the Slovak Research and Development Agency, grant APVV-0096-10, and by the Scientific Grant Agency of the Ministry of Education of the Slovak Republic and the Slovak Academy of Sciences, grants VEGA 2/0038/12, 2/0019/10.

{

}

\begin{thebibliography}{00}

\bibitem{Abramowitz}
Abramowitz M., Stegun I.~A.: Handbook of Mathematical Functions with Formulas, Graphs, and Mathematical Tables. \emph{National Bureau of Standards}, Tenth Edition, 1972.


\bibitem{Fisher}
Airey, J.~R., Irwin, J.~O., Fisher,  R.~A.:
Introduction to Tables of $Hh$ Functions.
\emph{British Association for the Advancement of Science, Mathematical Tables 1}, XXIV--XXXV, 1931. 

\bibitem{Benton}
Benton D., Krishnamoorthy K.: 
\emph{Computing discrete mixtures of continuous distributions: noncentral chisquare, noncentral $t$ and the distribution of the square of the sample multiple correlation coefficient}.
Computational Statistics \& Data Analysis \textbf{43} (2003), 249--267.

\bibitem{DISTEXPLORER}
Bristow, P.~A., Maddock, J.: DistExplorer: Statistical Distribution Explorer. \emph{Boost Software License}. Edition:  Version 1.0., 2012. \url{http://sourceforge.net/projects/distexplorer/}.

\bibitem{Guenther}
Guenther, W.~C.: 
\emph{Evaluation of probabilities for the noncentral distributions and the difference of two t variables with a desk calculator}. 
Journal of Statistical Computation and Simulation \textbf{6} (1978), 199--206.

\bibitem{Hahn}
Hahn, G.~J., Meeker, G.~J.:
Statistical Intervals: A Guide for Practitioners.  \emph{John Wiley \& Sons}, New York, 1991.

\bibitem{Holoborodko}
Holoborodko P.: Multiprecision Computing Toolbox for MATLAB.  \emph{Advanpix}, Yokohama. Edition: Version~3.4.3, 2013. \url{http://www.advanpix.com}.

\bibitem{Inglot}
Inglot, T.: \emph{Inequalities for quantiles of the chi-square distribution}.
Probability and Mathematical Statistics \textbf{30}  (2010), 339--351.

\bibitem{Ledwina}
Inglot, T., Ledwina, T.: \emph{Asymptotic optimality of a new adaptive test in regression model}.
 Annales de l'Institut Henri Poincar\'e \textbf{42} (2006), 579-–590.

\bibitem{Janiga2006}
Janiga, I., Garaj, I.:
\emph{OnE-sided tolerance factors of normal distributions with unknown mean and variability}.
{Measurement Science Review} \textbf{8} (2006), 12-–16. 

\bibitem{Johnson}
Johnson, N.~L., Kotz, S., Balakrishnan, N.:
Continuous Univariate Distributions, Volume 2. \emph{John Wiley \& Sons}, New York, Second Edition,  1995.

\bibitem{Kim}
Kim, J.:
{{Efficient Confidence Inteval Methodologies for the Noncentrality Parameters of the Noncentral $T$-Distributions.}
\emph{H.~Milton Stewart School of Industrial and Systems Engineering, Georgia Institute of Technology}, PhD Thesis, May 2007.

\bibitem{Krishnamoorthy}
Krishnamoorthy, K., Mathew, T.: 
Statistical Tolerance Regions: Theory, Applications, and Computation, \emph{John Wiley \& Sons}, New York, 2009.

\bibitem{Lenth}
Lenth, R.~V.:  \emph{Algorithm AS 243 — Cumulative distribution function of the non-central $t$ distribution}. 
Applied Statistics \textbf{38} (1989), 185--189.

\bibitem{boost}
Maddock, J., Bristow, P.~A., Holin, H., Zhang, X., Lalande, B., Rade, J., Sewani, G., van den Berg, T.,  Sobotta, B.: 
Noncentral $T$ Distribution. \emph{Boost C++ Libraries}, Edition: Version~1.53.0, 2012. \url{http://www.boost.org}.

\bibitem{Matlab}
The MathWorks Inc.:
MATLAB Edition: Version~8.0.0.783 (R2012b). Natick, Massachusetts, 2012. \url{http://www.mathworks.com}.

\bibitem{R}
R Development Core Team.: R: A Language and Environment for Statistical Computing. 
\emph{R Foundation for Statistical Computing}, Edition: Version~3.0.0, Vienna, Austria, 2013. \url{http://www.R-project.org}.

\bibitem{SAS}
SAS Institute Inc.: PROBT Function. SAS(R) 9.3 Functions and CALL Routines: Reference, 2013. \url{http://support.sas.com/}.

\bibitem{Student}
Student: \emph{The probable error of a mean}. Biometrika} \textbf{6} (1908), 1–-25.

\bibitem{Mathematica}
Wolfram Research, Inc.: 
Mathematica Edition: Version 9.0. \emph{Wolfram Research, Inc.}, Champaign, Illinois, 2013. \url{http://www.wolfram.com/mathematica/}.

\end{thebibliography}
\end{document}